\documentclass[journal,final]{IEEEtran}
\usepackage[noadjust]{cite}
\usepackage{color}

\ifCLASSINFOpdf
\else
\fi
\usepackage{threeparttable}

\usepackage{amsmath}
\usepackage{array}
\usepackage{graphicx}
\graphicspath{ {Figures/} }
\usepackage{multirow}

\begin{document}
\title{A Machine Hearing System for Robust Cough Detection Based on a High-Level Representation of Band-Specific Audio Features}

\author{Jes\'us Monge-\'Alvarez,
        Carlos Hoyos-Barcel\'o,
       Luis Miguel San-Jos\'e-Revuelta,
       and Pablo Casaseca-de-la-Higuera,~\IEEEmembership{Member,~IEEE}% <-this % stops a space
\thanks{Manuscript received March 16, 2018. Date of current version July 17, 2019. This work was supported by the Digital Health \& Care Institute Scotland as part of the Factory Research Project SmartCough/MacMasters. The authors would like to acknowledge support from University of the West of Scotland for partially funding C. Hoyos-Barcelo and J. Monge-Alvarez studentships. UWS acknowledges the financial support of NHS Research Scotland (NRS) through Edinburgh Clinical Research Facility. Acknowledgement is extended to Cancer Research UK for grant C59355/A22878. {\it (Corresponding author: Pablo Casaseca-de-la-Higuera}.}
\thanks{J. Monge-\'Alvarez and C. Hoyos-Barcel\'o are with the School of Computing, Engineering and Physical Sciences, University of the West of Scotland.}% <-this % stops a space
\thanks{L.M. San-Jos\'e-Revuelta is with Laboratorio de Procesado de Imagen, E.T.S.I. Telecomunicaci\'on, Universidad de Valladolid.}% <-this % stops a space
\thanks{P. Casaseca-de-la-Higuera is with the School of Computing, Engineering and Physical Sciences, University of the West of Scotland, Paisley Campus, High Street, Paisley, PA1 2BE, U.K. and also with Laboratorio de Procesado de Imagen, E.T.S.I. Telecomunicaci\'on, Universidad de Valladolid (e-mail: casaseca@lpi.tel.uva.es).}
\thanks{(c) 2018 IEEE. Personal use of this material is permitted. Permission from IEEE must be obtained for all other uses, in any current or future media, including reprinting/republishing this material for advertising or promotional purposes, creating new collective works, for resale or redistribution to servers or lists, or reuse of any copyrighted component of this work in other works.}
\thanks{\bf This is the final manuscript accepted and published (postprint) in IEEE TRANSACTIONS ON BIOMEDICAL ENGINEERING, VOL. 66, NO. 8, AUGUST 2019, pp. 2319-2330. Digital Object Identifier 10.1109/TBME.2018.2888998.}
}


\markboth{}%
{}



\maketitle

\begin{abstract}
Cough is a protective reflex conveying information
on the state of the respiratory system. Cough
assessment has been limited so far to subjective measurement
tools or uncomfortable (i.e., non-wearable) cough
monitors. This limits the potential of real-time cough monitoring
to improve respiratory care. Objective: This paper
presents a machine hearing system for audio-based robust
cough segmentation that can be easily deployed in mobile
scenarios. Methods: Cough detection is performed in
two steps. First, a short-term spectral feature set is separately
computed in five predefined frequency bands: $[0, 0.5)$,
$[0.5, 1)$, $[1, 1.5)$, $[1.5, 2)$, and $[2, 5.5125]$ kHz. Feature selection
and combination are then applied to make the short-term
feature set robust enough in different noisy scenarios. Second,
high-level data representation is achieved by computing
the mean and standard deviation of short-term descriptors
in 300 ms long-term frames. Finally, cough detection is
carried out using a support vector machine trained with data
from different noisy scenarios. The system is evaluated using
a patient signal database which emulates three real-life
scenarios in terms of noise content. Results: The system
achieves 92.71\% sensitivity, 88.58\% specificity, and 90.69\%
Area Under Receiver Operating Characteristic (ROC) curve
(AUC), outperforming state-of-the-art methods. Conclusion:
Our research outcome paves the way to create a device for
cough monitoring in real-life situations. Significance: Our
proposal is aligned with a more comfortable and less disruptive
patient monitoring, with benefits for patients (allows
self-monitoring of cough symptoms), practitioners (e.g., assessment of treatments or better clinical understanding of
cough patterns), and national health systems (by reducing
hospitalizations).
\end{abstract}

\begin{IEEEkeywords}
Cough detection, machine hearing, respiratory care, patient monitoring, spectral features.
\end{IEEEkeywords}

% \fi
\IEEEpeerreviewmaketitle

% SECTION I: INTRODUCTION
\section{Introduction}
\IEEEPARstart{C}{ough} is a protective reflex with a characteristic sound and associated body movement. It is associated with over one hundred pathological conditions, and it is therefore one of the main causes for patients seeking medical care. Many of these pathological conditions are related to the respiratory system (e.g., chronic obstructive pulmonary disease (COPD) or asthma), while others are seasonal diseases like influenza, allergies or cold \cite{Fontana2007}. Additionally, cough can be related to lifestyle (smokers, sedentary people) or certain physical activities (athletes) \cite{Hull2017}.

Even though it is a frequent symptom, there is no clear consensus on the definition of cough \cite{Chung2009}. The European Respiratory Society Task Force \cite{ERSGuidelines} provides the following: 'A forced expiratory manoeuvre, usually against a closed glottis and associated with a characteristic sound'. Similarly, there is lack of standardisation in the methods to assess cough. There exist objective and subjective methods to assess cough, and both have their counterparts \cite{Birring2015}.

Subjective methods are based on diaries or quality-of-life questionnaires where patients can provide their appreciation of cough severity \cite{French2002}. One the one hand, these methods are cheap and readily applicable in primary care but, on the other hand, they might be biased due to the physical and psychological comorbidity of cough (e.g., incontinence, chest pain or social embarrassment), inter-expert variability \cite{Chung2006} and other factors such as personality or mood \cite{Brignall2008}.

The development of digital technologies has fostered the emergence of healthcare devices to objectively monitor cough. The operation of these devices relies on pattern recognition engines primarily based on features extracted from cough sounds and complementary signals like electromyography of chest movement. However, many of these systems have only been tested in controlled environments where patients did not perform any movement or physical activity \cite{Smith2008}, \cite{Amrulloh2015} or force the users to wear complex recording systems \cite{Smith2008}, \cite{Coyle2005}. A portable cough monitor relying on audio recordings and able to cope with noisy real-life environments could be implemented on a smartphone for continuous real-time monitoring of respiratory patients. This would constitute a reliable piece of technology for practitioners so that the potential of telehealth in the context of respiratory disease could be leveraged. In addition, from the patient's point of view, this monitoring system would be more comfortable and bring minimal disruption to their daily activities. This way, they would be less conciousness of their medicalisation.

\begin{figure*} % [b]
\centering
\includegraphics[width=14cm]{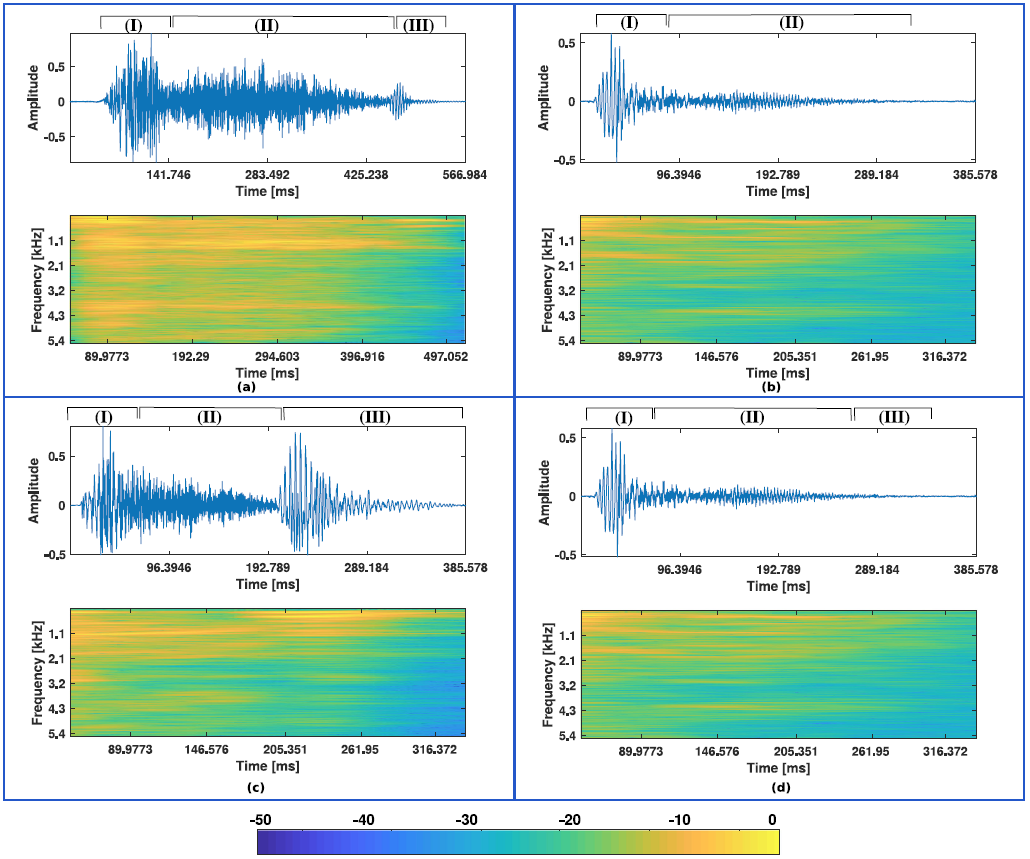}
\caption{Representation of different cough events and their spectrograms where the specific phases have been detailed: (I) explosive phase, (II) intermediate phase, and (III) voiced phase. Events: (a) strong intermediate phase; (b) absent vocal phase; (c) strong vocal; (d) weak intermediate and vocal.}
\label{Fig1}
\end{figure*}

Audio cough events are non-stationary signals composed of three phases: explosive, intermediate and voiced. These signals do not have a clear formant structure and are characterised by a sparse spectral content. The average length of cough events has been reported to be around 300 ms \cite{Smith2007}. Fig.  \ref{Fig1} shows four cough events with different features: strong intermediate phase (top-left), absent vocal phase (top-right), strong vocal phase (bottom-left), and weak intermediate and vocal phases (bottomright).
Even though cough events present a similar waveform, there is inter- and intra-patient variability affecting both the duration and intensity of the three phases.

Most studies aiming at cough segmentation are based on the primary approach of machine hearing \cite{IntAudioAna} (the set of signal processing and machine learning techniques for audio signal analysis). This treats the audio signal as being linear and stationary for short intervals of time (between 20 and 100 ms). We will refer to this approach hereafter as short-term. Matos \textit{et al.}  \cite{Matos2006}, used a combination of MFCC (Mel Frequency Cepstral Coefficients) and hidden Markov models to achieve an average 82\% cough detection accuracy. You \textit{et al.} \cite{You2017} used an ensemble of multiple frequency subband features. The classification was based on a linear support vector machine (SVM). Recall values around 74\% were reported on real data (classification of each subband separately) with an overall 82\% performance after integration. Amrulloh \textit{et al.} \cite{Amrulloh2015} employed MFCC together with entropy and non-gaussianity measures for cough segmentation in pediatric wards. They used an artificial neural network for classification, achieving 93\% sensitivity (SEN). However, these high figures were reported in a quite environment. The work in \cite{Sert2015} used MFCC and a SVM to recognise cough events among other sounds like throat clearings, speech or knockings within an office life environment. 63.6\% SEN was only reported for cough events.

Other proposals have achieved promising detection figures in real-life noisy scenarios. Amoh and Odame  \cite{CNN2016} employed convolutional neural networks (CNN) and a recurrent neural network (RNN) to perform cough segmentation using time-frequency representation of audio frames obtained from a lapel microphone. Both networks offered SEN around 83\%, whereas the specificity (SPE) obtained from the CNN was better (93\%) than for RNN (75\%). Even though the audio signal was recorded during real-life activity, the quality of the acquired signal was favoured by the proximity of the lapel microphone to the mouth. Our work in \cite{Monge2018} achieved robust segmentation of audio cough events using short term processing. The proposal applied moment theory to characterise adjacent frames and frequency bands of the \textit{cepstrogram} audio signal representation. A \textit{k}-Nearest Neighbour classifier provided SEN and SPE values above 85\%. The experimental set up included an artificial scenario where the signals were contaminated with noise at different Signal to Noise Ratios ranging from -6 dB to 15 dB. Smartphone-recorded data in different noisy scenarios (including when the device was carried in a pocket or bag) was used to validate the method. This work also demonstrated that classical feature sets such as the MFCC employed in \cite{Matos2006, Amrulloh2015, Sert2015} failed to perform on challenging noisy environments.

The short-time approach employed in \cite{Matos2006, You2017, Amrulloh2015, Sert2015,CNN2016,Monge2018} gives a simplistic and time-affordable analysis, enabling the classification of signal frames as belonging to a cough event or not. However, when a high-level representation of the data is used, the actual segmentation of cough events can improve while keeping the classification scheme simple. Besides, this representation also favours system robustness \cite{Foggia2015}, \cite{Grzeszick2017}. In this paper we propose a machine hearing system for robust cough segmentation based on a high-level data representation. The proposed method first computes a number of short-term features in relevant frequency bands specific to the audio-cough spectrum. The most meaningful features are then selected and combined in a high-level representation to perform robust cough detection in noisy conditions. Results on real patient data show that the proposed approach overcomes the best performing of the recently proposed robust cough detectors \cite{You2017,CNN2016,Monge2018}

The remaining of the paper is organised as follows: Section \ref{SectionII} constitutes the materials section and presents the patient signal database used for the evaluation of the system. Section \ref{SectionIII} (methods) describes the proposed machine hearing system for robust cough detection. Results are presented in Section \ref{SectionIV} and discussed in Section \ref{SectionV}. Finally, Section \ref{SectionVI} summarizes the main conclusions of the paper.

% SECTION II: PATIENT SIGNAL DATABASE
\section{Materials} \label{SectionII}
Ambulatory recordings from thirteen adult patients acquired at the Outpatient Chest Clinic, Royal Infirmary of Edinburgh (UK) – all of them presenting cough as a symptom of their underlying condition (see Table \ref{TableI}) – constitute the information source of the present study. The study was carried out in accordance with the Declaration of Helsinki and was approved by the NHS Lothian Research Ethics Committee (REC number: 15/SS/0234). Subjects provided their informed consent before the recordings. The acquisition protocol is described below.

The first part simulates a low-noise environment. In this situation, the patient, who is sitting on a chair, is requested to speak or read aloud. From time to time, we asked the patient to produce other non-cough events such as throat clearing, swallowing (by drinking a glass of water), blowing nose, sneezing, breathless breathing or laugh (by reading a joke or a humour comic).

The second part of the protocol emulates a noisy environment with an external source of contamination (the patient does not produce the noisy background sounds). To do so, we repeated the experiment in part one with either a television set or radio player on, and also allowing noise from the hospital corridor being recorded as well (e.g.: babble noise, typing noise or a trolley in movement). This second part is a moderately noisy environment.

Finally, the third part of the protocol was designed to represent noisy environments where the own patients also become a source of contamination because of their movements and other activities. In this case, the patient could freely move around the room while we asked her to carry out some activities such as opening/closing the window, opening/closing a drawer, moving a chair, washing hands, lying on the bed and standing up immediately, typing, putting their coat on and taking it off immediately after that, picking up an object from the floor, among others. Similar to the second part, a TV or radio was on and the door was left open to record corridor noisy sounds as well. Equally, while the patient was performing these activities, we requested her to produce other non-cough foreground events as in the first and second part. The third part thus represents a high noise scenario.

Every single part of the protocol lasted twenty minutes, so a total of thirteen hours of recording were acquired. The digital recorder was placed on a table in the centre of the room. All the recorded cough events were spontaneous. The number of cough and non-cough events for each patient and part of the protocol
is also presented in Table \ref{TableI}. Signals were recorded in \textit{wav} format using a Samsung S6 Edge smartphone, at 44.1 kHz sampling frequency, with 16 bits per sample. The recording app was configured to ignore sounds 70 dB below the maximum dynamic range of the device. Audio files were manually annotated
on a time-frame basis. If a frame contained samples belonging to cough and non-cough events, the class contaning the majority of samples was selected. The acquisition protocol is similar to the one used in other studies \cite{Matos2006,You2017,Sert2015} although it is more diverse -- in terms of types of noisy sounds -- and presents a higher degree of contamination than the ones used in \cite{Amrulloh2015,CNN2016}.

\begin{table*} % [t]
\centering
\caption{Clinical information of the patient population and distribution of cough and non-cough events.}
\label{TableI}
	\begin{tabular}{c|c c c|c c c| c c c|}
		\hline \hline
        \textbf{ID} & \textbf{Age} & \textbf{Gender} & \textbf{Medical condition}    &      \multicolumn{3}{c|}{No. of cough events}    &    \multicolumn{3}{c|}{No. of non-cough events}     \\
          &   &  &  &  Part 1  &   Part 2  & Part 3 & Part 1 &  Part 2  &  Part 3 \\
        \hline
        1 	& 70 & Female 	& Bronchiectasis & 21 & 16 & 37 & 223 &410 & 246   \\
        2 	& 45 & Male 	& Asthma   & 28 & 36 & 32 & 243 & 512 & 346\\
        3 	& 69 & Female 	& COPD     & 37 & 27 & 39 & 478 & 370 & 522 \\
        4 	& 48 & Male 	& COPD    & 28 & 15 & 43 & 288 & 105 & 245 \\
        5 	& 48 & Female 	& Bronchiectasis  & 26 & 27 & 47 & 210 & 182 & 269  \\
        6 	& 72 & Female 	& Asthma & 33 & 34 & 30 & 170 & 197 & 197 \\
        7 	& 66 & Female 	& COPD   & 19 & 11 & 18 & 188 & 374 & 247 \\
        8 	& 66 & Female 	& Bronchiectasis & 17 & 12 & 21 & 164 & 256 & 230 \\
        9 	& 61 & Female 	& COPD    & 45 & 39 & 50 & 403 & 329 & 376 \\
        10	& 68 & Female 	& Bronchiectasis  & 45 & 33 & 32 & 414 & 365 & 319 \\
        11 	& 65 & Female 	& COPD    & 26 & 27 & 17 & 166 & 242 & 200 \\
        12 	& 72 & Female 	& Asthma & 86 & 84 & 74 & 366 & 552 & 704 \\
        13 	& 67 & Male 	& COPD   & 37 & 32 & 28 & 173 & 195 & 151 \\
        \hline \hline
	\end{tabular}
\end{table*}

\begin{figure} % [b]
\centering
\includegraphics[width=\columnwidth]{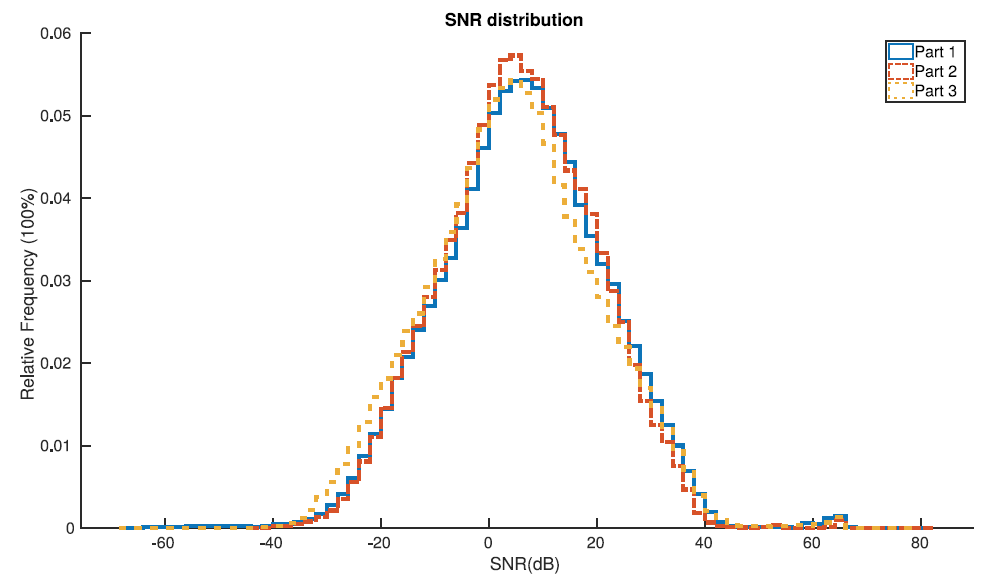}
\caption{Patient-aggregated SNR distribution for each part of the protocol}
\label{Fig2new}
\end{figure}

\begin{table}
\centering
\caption{Patient-specific SNR statistics}   %%%%%%%%%% HACEMOS LA TABLA COMO UNA FIGURA DENTRO DE LA TABLA
\label{TableII}
	\begin{tabular}{c}
		\includegraphics[width=\columnwidth]{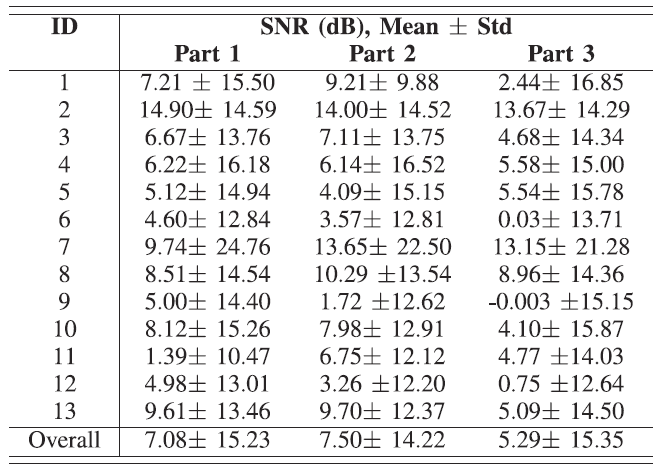}
	\end{tabular}
\end{table}

Fig. 2 shows the patient-aggregated SNR distribution for each part of the protocol. In order to compute SNR values, we substracted the surrounding noise power to the power of each annotated cough frame and divided the result by the noise power. To obtain the latter, the average power of the preceding and following non-cough frames was computed. A higher concentration of low SNRs can be observed for the third part in the figure. SNR mean and standard deviation are presented for each patient in Table II.

% SECTION III: METHODS
\section{Methods} \label{SectionIII}

\begin{figure*}
	\centering
	\includegraphics[width=\textwidth]{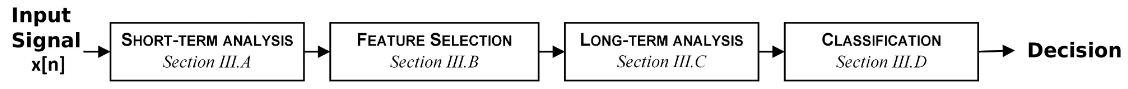}
	\caption{Processing pipeline of the proposed cough detection system with specific references to the  sections describing each block.}
	\label{pipeline}
\end{figure*}

The processing pipeline of the overall system is depicted in Fig. \ref{pipeline}. Each block is described in the following subsections. The input signal is downsampled at 11.025 kHz and split in 75 ms frames with 19s overlap to control boundary effects. The spectrogram of each frame is computed to extract spectral short-term features which are further processed to obtain a high-level representation after feature selection. The use of 75 ms frames is justified on the basis of the need of an accurate spectral estimation while accounting for the non-stationarity nature of the signals. It also suits the distribution of the three different phases in a cough event. The explosive phase usually spans the first 25\% of the event, the intermediate one, the second and third quarter, and the vocal one, the last 25\%. The obtained high-level features are used afterwards to feed a SVM for final classification.

% SUB-SECTION III.A: SHORT-TERM DESCRIPTORS
\subsection{Short-term descriptors} \label{SectionIII.A}

% SUB-SECTION III.A.1: FREQUENCY BANDS-BASED UNIDIMENSIONAL SPECTRAL FEATURES
\subsubsection{Band-based unidimensional spectral features} \label{SectionIII.A.1}
Spectral features are commonly used to characterise audio \cite{Ramalingam2006} and biomedical signals \cite{Poza2008}. Thus, they constitute a sensible option to identify cough patterns. The main advantage of these features is their low computational complexity once the spectrum has been obtained. Besides, some of them have a physical interpretation. Even though they are usually computed over the whole signal spectrum, we followed a subband-based approach in this work after observing typical spectral patterns of coughs obtained from patients, control subjects (smokers and non-smokers), children and babies.

\begin{figure}[b] % [b]
\centering
\includegraphics[width=\columnwidth]{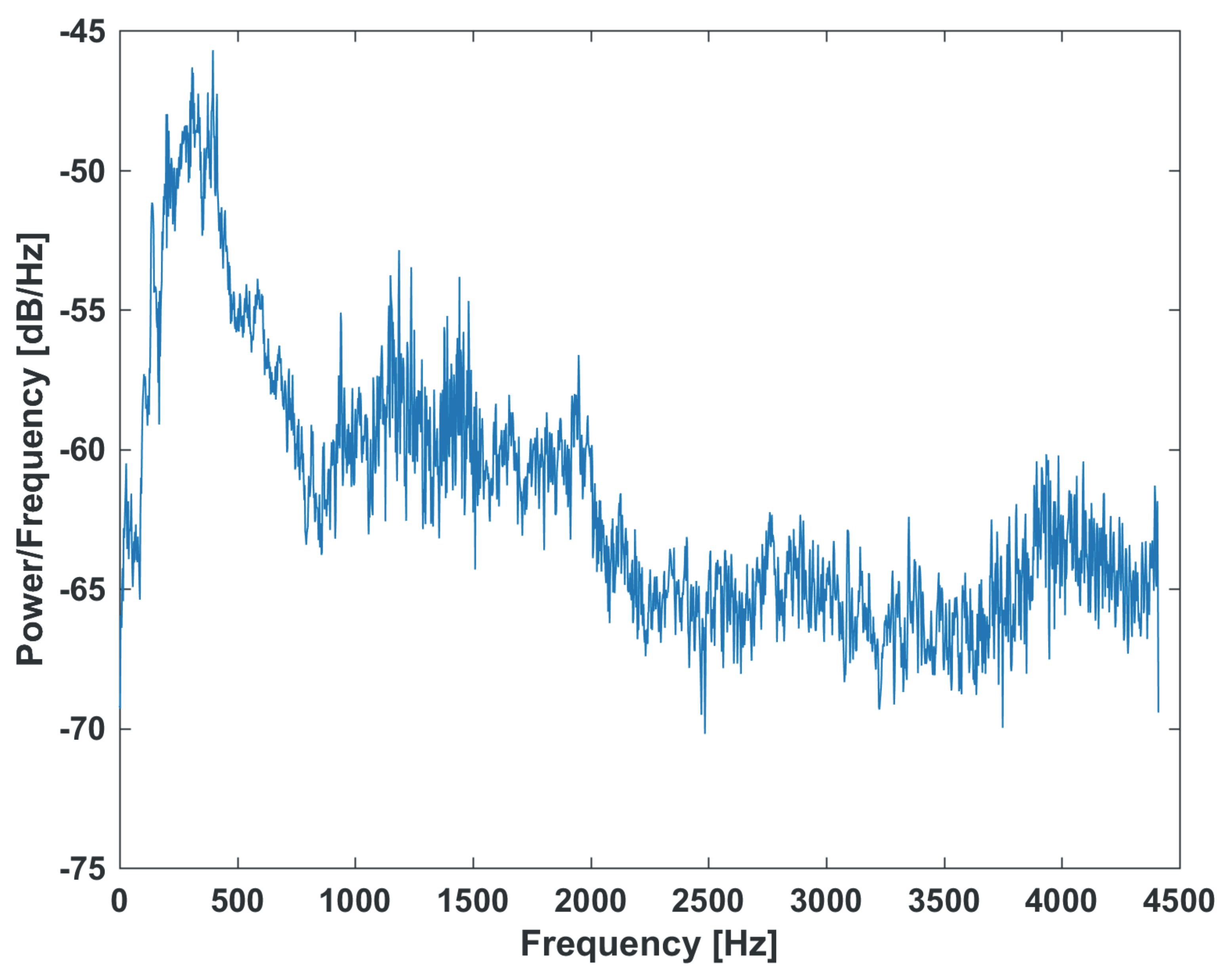}
\caption{Sample average periodogram of the recorded cough events.}
\label{Fig2}
\end{figure}

The average periodogram of cough events from these subjects was computed to account for intra- and inter-person variability \cite{Smith2007}. This periodogram showed a prominent peak around 500 Hz and secondary peaks between 1000 and 1500 Hz (see Fig \ref{Fig2}).

Five frequency bands were defined: \([0,0.5), [0.5,1), [1,1.5), [1.5,2), [2,5.5125]\) kHz. We hold the hypothesis that features aiming at identifying dominant frequencies such as centroid or crest factor (see description below) will be more helpful to characterise the first and the third frequency bands, for instance, while other like flatness or entropy measures will equivalently perform in the second, fourth and fifth band since those bands do not have prominent peaks. This way, a fine-grained characterisation of cough patterns is achieved with better representation than the one obtained from features computed over the whole spectrum.

Each of the following features is computed for every frequency band referred above. For frequency decomposition, the one-sided Welch's power spectral density (PSD) \cite{Haykin2009} of each 75 ms frame is calculated using three sub-frames of 275 samples with no overlap. Henceforth, the index \textit{j} refers to the frequency band: \(j=1 \rightarrow [0,0.5)\) kHz, \(\cdots\), \(j=5 \rightarrow [2,5.5125)\) kHz. \(PSD_j[k]\) and \(f_j[k]\) respectively represent the corresponding part of the Welch's PSD and the vector of discrete frequencies in the band, respectively.

\textbf{Spectral centroid}, which can be understood as the spectral centre of gravity \cite{Ramalingam2006}:
\begin{equation}
\label{Eq1}
SpecCent(j)=\sum_{k}^{}f_j[k] \cdot PSD_j[k]/\sum_{k}^{}PSD_j[k]
\end{equation}

\textbf{Spectral bandwidth}, a measure of the spectral distribution \cite{Ramalingam2006}:
\begin{equation}
\label{Eq2}
SpecBand(j)= \frac{\sum_{k}^{} (f_j[k]-SpecCent(j))^2 \cdot PSD_j[k]}{\sum_{k}^{}PSD_j[k]}
\end{equation}

\textbf{Spectral Crest Factor}, this feature detects the dominant frequency of the spectrum \cite{Ramalingam2006}:
\begin{equation}
\label{Eq3}
C = 1/(\max\{f_j[k]\}-\min\{f_j[k]\}+1)
\end{equation}
\begin{equation}
\label{Eq4}
SpecCrestFac(j) = \frac{max\{PSD_j[k]\}}{C \cdot \sum_{k}^{}PSD_j[k]}
\end{equation}

\textbf{Spectral flatness}, a high value means a white-noise-like spectrum, with flat spectral content \cite{Ramalingam2006}:
\begin{equation}
\label{Eq5}
SpecFlat(j) = \frac{\exp(E\{\log(PSD_j[k])\})}{E\{PSD_j[k]\}}
\end{equation}
 where \(E\{\cdot\}\) refers to the expected value operator.

\textbf{Spectral flux}, which accounts for abrupt spectral changes between adjacent frames \cite{AudioMatlabChap4}:
\begin{equation}
\label{Eq6}
SpecFlux(j) = \sum_{k}^{} (PSD_j^i[k]-PSD_j^{i-1}[k])^2
\end{equation}
\(PSD_j^i[k]\) refers to the PSD calculated over the \textit{i-th} frame.

The \textbf{spectral roll-off} is defined as the frequency at which 85\% of the energy is included \cite{AudioMatlabChap4}:
\begin{equation}
\label{Eq7}
\sum_k^{k85_j}PSD_j[k] = 0.85 \cdot \sum_k PSD_j[k]
\end{equation}
\begin{equation}
\label{Eq8}
SpecRollOff(j) = f[k85_j]
\end{equation}
where $k85_j$ is the $k$th-value in frequency factor $f_j[k]$ below which 85\% of the total energy is included.

\textbf{Ratio f50 vs f90}, defined as the ratio between the frequencies at which 50\% and 90\% of the energy is included \cite{Wisniewski2011}:
\begin{equation}
\label{Eq9}
\sum_k^{k50_j} PSD_j[k] = 0.5 \cdot \sum_k PSD_j[k]
\end{equation}
\begin{equation}
\label{Eq10}
\sum_k^{k90_j} PSD_j[k] = 0.9 \cdot \sum_k PSD_j[k]
\end{equation}
\begin{equation}
\label{Eq11}
f50f90Ratio(j) = f[k50_j] / f[k90_j]
\end{equation}

The \textbf{spectral peak entropy} is an entropy measure based on the peaks and valleys of the spectrum. First, the local maxima (\textit{lm}) of the spectrum are sought to compute \cite{Wisniewski2011}:
\begin{equation}
\label{Eq12}
P_j[k_{lm}] = PSD_j[k_{lm}] / \sum PSD_j[k_{lm}]
\end{equation}
The term \(k_{lm}\) refers to the discrete frequencies at which the local maximum are found.
\begin{equation}
\label{Eq13}
SpecPeakEn(j) = -1 \cdot \sum P_j[k_{lm}] \cdot \log_{10}(P_j[k_{lm}])
\end{equation}

\textbf{Spectral Renyi entropy}, a generalised measure of uncertainty or randomness \cite{Poza2008}:
\begin{equation}
\label{Eq14}
SpecRenyiEn(j) = 1/(1-q) \cdot \log(\sum_k^{}PSD_j[k])^q
\end{equation}
where \(q=4\) was used for this work.

\textbf{Spectral kurtosis}, a descriptor of the spectral shape.
\begin{equation}
\label{Eq15}
\mu_j = E\{PSD_j[k]\}
\end{equation}
\begin{equation}
\label{Eq16}
\sigma_j = \sqrt[2]{E\{(PSD_j[k]-\mu_j)^2\}}
\end{equation}
\begin{equation}
\label{Eq17}
SpecKurt(j) = E\{((PSD_j[k]-\mu_j)/\sigma_j)^4\}
\end{equation}

\textbf{Spectral skewness}, a statistical measure of the spectral asymmetry:
\begin{equation}
\label{Eq18}
SpecSkew(j) = E\{((PSD_j[k]-\mu_j)/\sigma_j)^3\}
\end{equation}

The \textbf{relative power} is the ratio between the power at each frequency band and the total power in the frame:
\begin{equation}
\label{Eq19}
RP(j) = \sum_k PSD_j[k] / \sum_k PSD[k]
\end{equation}
where \(PSD[k]\) is the complete Welch's PSD of the frame.

Finally, the \textbf{spectral entropy} is the entropy measure of the relative power \cite{AudioMatlabChap4}:
\begin{equation}
\label{Eq20}
SpecEn = -1 \cdot \sum RP(j) \cdot \log_2(RP(j))
\end{equation}

% SUB-SECTION III.A.2: OTHER AUDIO FEATURES
\subsubsection{Other audio features} \label{SectionIII_A_2}
The features presented in Section \ref{SectionIII.A.1} were complemented by other typical audio features summarised in Table \ref{TableIII}.

\begin{table} % [t]
\centering
\begin{threeparttable}
\caption{Description of other short-term features}
\label{TableIII}
	\begin{tabular}{m{13mm}|c m{22mm} c}
		\hline \hline
        \textbf{Feature} & \textbf{Algorithm} & \textbf{Parameters} & \textbf{Dimension} \\
        \hline
        HR \tnote{1} 	& \cite{AudioMatlabChap4} 	&  \centering---							 & 1 \\
        \hline
        Root MFCC 		& \cite{Sharan2016} 		& * \# filters: 30 \newline
        											  * [0,4000] Hz \newline
                                                      * Root value: 1/2 \newline
                                                      * \(2^{nd}-14^{th}\) DCT 			 			 & 13 \\
        \hline
        ASF \tnote{2} & \cite{MPEG7Chap2} & * [62.5,4000] Hz 							 & 13 \\
        \hline
        NASE \tnote{3} & \cite{MPEG7Chap2} &  * [62.5,4000] Hz 			& 14 \\
        \hline
        TI \tnote{4} 		& \cite{WisniewskiJBHI2015} & \centering--- & 1 \\
        \hline
        ChroEn \tnote{5} 	& \cite{AudioMatlabChap4} 		& \centering--- & 1 \\
        \hline
        SSCH \tnote{6} & \cite{SSCH2006} & 	* \# filters: 30 \newline
        															* 3 Barks width \newline
                                                                    * $[0,4000]$ Hz \newline
                                                                    * \# bins: 38 \newline
                                                                    * \(2^{nd}-14^{th}\) DCT 		 & 13 \\
        \hline \hline
	\end{tabular}
    \begin{tablenotes}
     \item[1] Harmonic Ratio (HR)
     \item[2] Audio Spectrum Flatness (ASF)
     \item[3] Normalized Audio Spectrum Envelope (NASE)
     \item[4] Tonal Index (TI)
     \item[5] Chromatic Entropy (ChroEn)
     \item[6] Subband Spectral Centroid Histograms (SSCH)
   \end{tablenotes}
   \end{threeparttable}
\end{table}

\subsubsection{Justification of the Selection of Short-Term Features} 
The selection of spectral-shape features described in Section III-A1 is justified on the basis of both their simplicity (which contributes to the efficiency of the system) and the fact that they constitute physical acoustic features, since there is no assumption about the data such as the presence of prominent spectral peaks. Besides, they are commonly used in machine hearing and in the analysis of other signals in biomedical applications. This makes them especially suitable to the nature of our problem.

The rest of employed feature sets (see Section III-A2) have also been used in both types of applications. For instance, TI has been employed to analyse asthma wheezes [28]. SSCH [29], root MFCC [26], and HR [24] have been applied to robust speech detection. Since speech signals are usually interleaved with cough patterns, their incorporation seems sensible. NASE and ASF [27] are part of the MPEG-7 standard and as such they are interesting for the analysis of multimedia sounds. Similarly, ChroEn has been used for music analysis [24]. All these features have been incorporated into the analysis so as to cover a wide range of sounds. Apart from the physical audio features mentioned in the previous paragraph, we have also tried to ensure that perceptual features based in different scales such as Mel (MFCC and TI), Octaves (NASE, ASF, and ChroEn) or Bark (SSCH) are considered. Notice that spectral analysis involving these features has been limited to the $0-4000$ Hz range, since most descriptors are specifically designed for speech recognition in this band. In addition, when filter banks are involved, limiting the frequency range keeps the number of filters ($\sharp$ filters in Table III) bounded.

% SUB-SECTION III.B: FEATURE SELECTION
\subsection{Feature selection} \label{SectionIII.B}
The short-term feature set described in the preceding section led to an overall dimension of \((12 (\text{spectral features}) \cdot 5 (\text {bands}) + 1 ({SpecEn})) + 56 (\text{Table \ref{TableIII}}) = 117\) features. Feature selection is thus necessary to improve efficiency. In addition, by removing redundant information, classification performance is expected to improve, also avoiding the curse of dimensionality \cite{Dougherty2013}.

An extra difficulty at the time of carrying out feature selection in this study is to find the most relevant feature set regardless the ambient noise which, in our study, was different for each part of the acquisition protocol. The following selection approach was adopted to cope with this problem:

\begin{enumerate}[\setlength{\IEEElabelindent}{0pt}]
\item 10\% of the observations of the feature space were randomly selected for each part of the protocol. The class ratio was kept unaltered in the selected partition.
\item Their intrinsic dimension was estimated using a maximum likelihood estimator (MLE) \cite{Levina2005}. All the obtained values ranged from 25 to 30.
\item The Relieff algorithm \cite{Relieff} - a widely used supervised feature selection algorithm for two-class problems - was applied to the selected observations in order to identify the best 29 features in each part.
\item The following combination procedure was applied step by step to select the best 29 features among the three sets obtained in each part of the protocol. To build the final set, each step is followed in order. The process finishes once 29 features are selected (i.e., if 29 features are selected after the $i$-th step, steps $\{i+1,\ldots,7\}$ are not followed). For each step, features are selected with the following criteria:
	\begin{enumerate}
	\item[\(1^{st}\))] features which belong simultaneously to the best thirty features in all the three parts (henceforth, regardless the Relieff ranking index within each part).
    \item[\(2^{nd}\))] features belonging to the best thirty features in the second and the third part
    \item[\(3^{rd}\))] features belonging to the best thirty features in the first and the third part
    \item[\(4^{th}\))] features belonging to the best thirty features in the first and the second part
    \item[\(5^{th}\))] features only selected in the third part
    \item[\(6^{th}\))] features only selected in the second part
    \item[\(7^{th}\))] features only selected in the first part
	\end{enumerate}
\end{enumerate}

\begin{figure*}
\centering
\includegraphics[width = 0.7\textwidth]{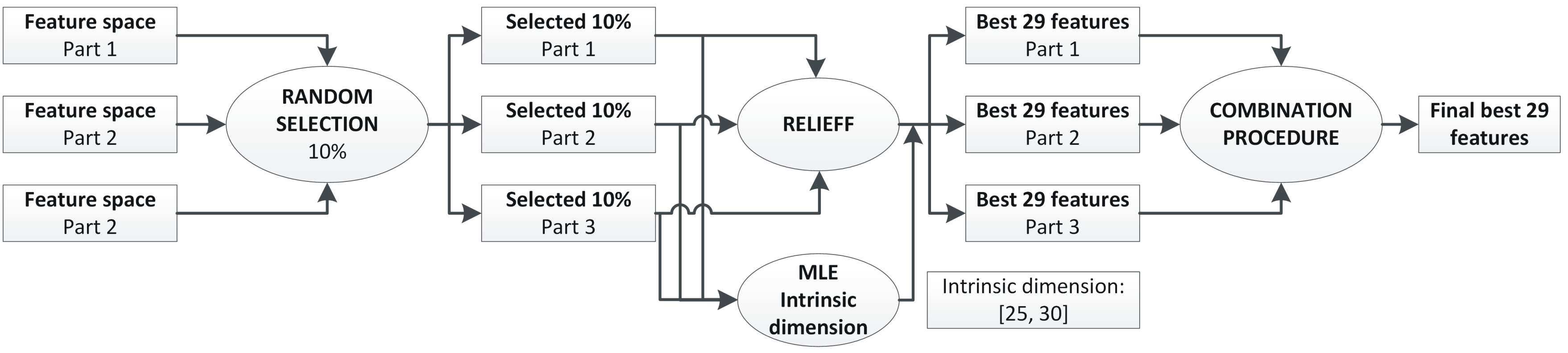}
\caption{Pipeline of the the feature selection process.}
\label{Fig3}
\end{figure*}

This method is based on the assumption that if a feature is a good descriptor in noisy environments, it will also be in more favourable conditions. Fig. \ref{Fig3} summarises the described procedure, for the sake of clarity.

% SUB-SECTION III.C: HIGH-LEVEL DATA REPRESENTATION
\subsection{High-level data representation} \label{SectionIII.C}
Even though a cough event lasts 300 ms on average, their inter-event length distribution is variable. A cough episode may content from two to dozens of cough events. Therefore, if a long-term scale longer than 300 ms is selected, there is a high risk that the system misclassifies isolated cough events. We thus selected the long-term frames for our method as composed of five short-term frames  with an overlap of one short-term frame. This yields an effective duration of \((75-19) \cdot 4) + 75 = 299\) ms. For cough events longer than 300 ms, there is still a risk of
identifying them as different consecutive cough events. However, this can easily be dealt with at postprocessing by grouping consecutively detected coughs as belonging to the same event.

\begin{figure*}
\centering
\includegraphics[width = 0.8\textwidth]{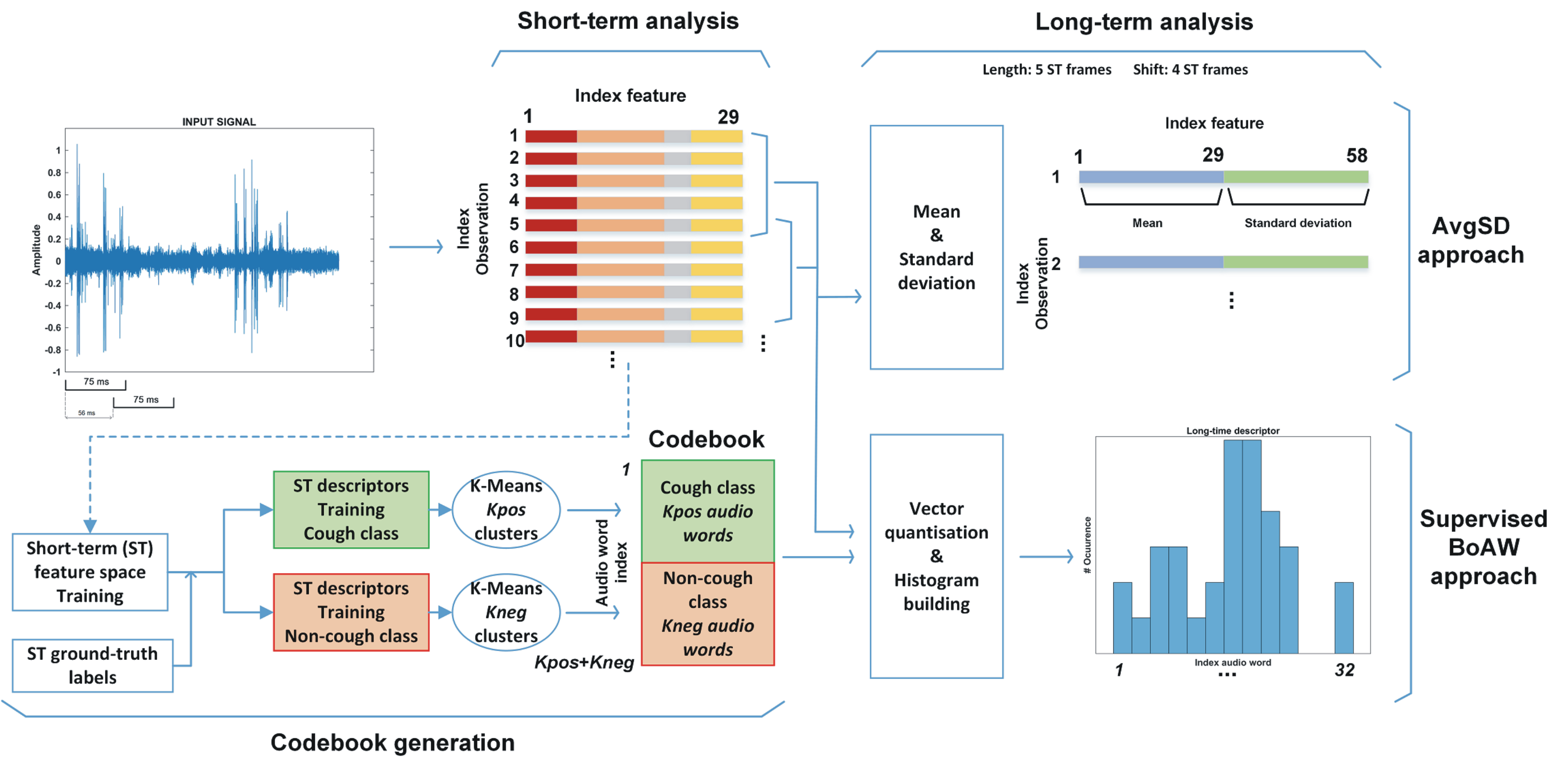}
\caption{Explanatory diagram of how high-level representation is obtained from short-term features. In the AvgSD method, feature-wise mean and standard deviation are concatenated. In the BoAW approach, 16 audio words for each class (\(K_{pos} = K_{neg} = 16\)) are generated using K-Means.}
\label{Fig4}
\end{figure*}

We evaluated two methods to obtain a high-level representation of the feature space:
\begin{itemize}[\setlength{\IEEElabelindent}{0pt}]
\item \textbf{Mean and standard deviation (referred to as AvgSD hereinafter):} this is the baseline representation \cite{AudioMatlabChap4}. Each long-term observation concatenates the feature-wise average and standard deviation of the corresponding short-term frames. Therefore, the dimension of the long-term feature space is twice the short-term one.

\item \textbf{Supervised BoAW:} this paradigm was adopted for audio signal processing from the well-established techniques used to process text (bag-of-words) and images (bag-of-visual-words). It has been used for song retrieval \cite{Riley2008}, multimedia event detection \cite{Pancoast2012} or robust detection of audio events \cite{Foggia2015}, for example. The rationale behind BoAW is that the audio stream can be divided into small perceptual units of hearing, the so-called audio words. The distribution of these audio words over long-time intervals allows characterising different sounds events. The codebook or dictionary of audio words is generated using a clustering algorithm, where each word corresponds to a cluster centroid. The codebook is then used in a vector quantisation step to replace each
    short term feature vector with the closest audio word. Finally, a histogram is built by counting the number of occurrences of each audio word over a long-time frame. This histogram constitutes the final feature vector to characterise the audio event in the corresponding long-term frame \cite{Plinge2014,Foggia2015}.\\
In the supervised version of BoAW, the training group is divided based on the ground-truth labels \cite{Plinge2014}. Later, the clustering algorithm (K-Means in our study \cite{KMeans}) is applied to the class-separated training sets to generate their audio words. The final codebook is composed by joining the audio words of each class-separated training set.

Fig. \ref{Fig4} shows an schematic procedure of how high-level representations are obtained from short-term features.
\end{itemize}

% SUB-SECTION III.D: CLASSIFICATION
\subsection{Classification} \label{SectionIII.D}
The machine hearing system aims to discriminate between audio-cough events and non-cough events regardless the superimposed noisy background sounds. This is posed as a two-class pattern classification problem, where cough is the positive class, and any non-cough sounds belong to the negative one.

\begin{figure}
\centering
\includegraphics[width = \columnwidth]{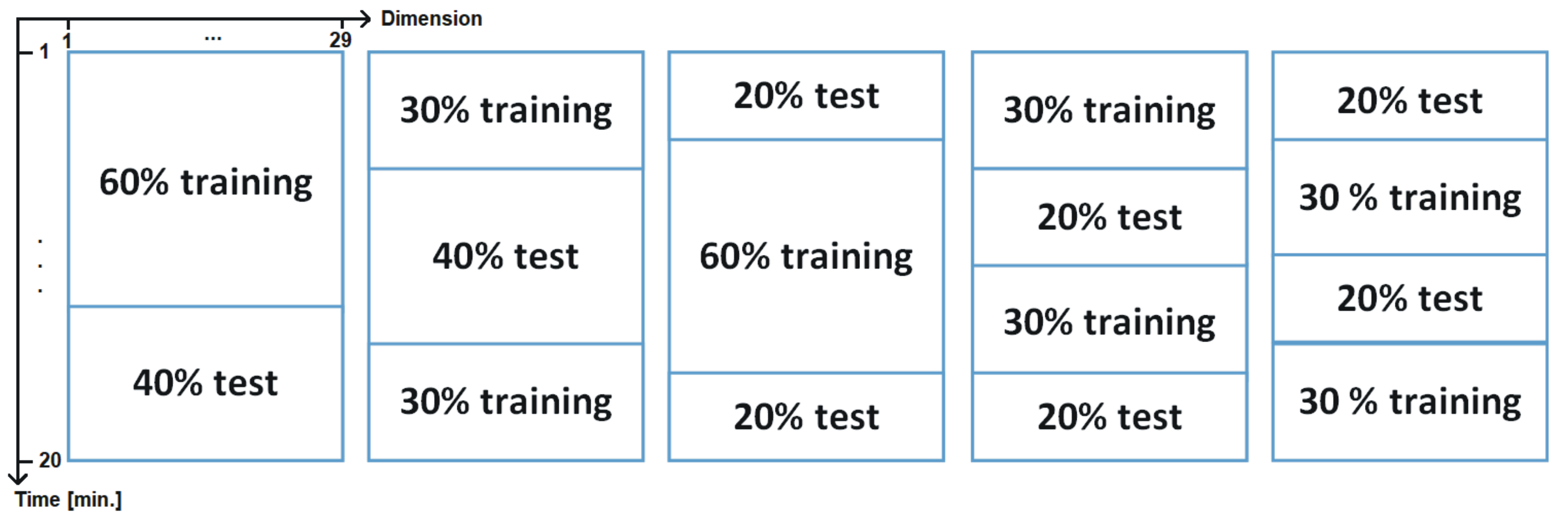}
\caption{Representation of the five block-division train-test partitions employed in the study.}
\label{Fig5}
\end{figure}

We employed SVMs with a 2nd order polynomial kernel for the classification step. The long-term approach used in this work imposes keeping the temporal alignment of the short-term descriptors, so five block-division partitions -- depicted in Fig. \ref{Fig5} -- are used. The feature space for each patient is thus divided based on these partitions. The final training and test groups are built by joining the corresponding training and test blocks for all patients. The definition of train and test sets constitutes a 5-fold cross-validation process, where blocks have been predefined to ensure that test patterns are not close to training patterns.

SEN, SPE and Area Under the Receiver Operating Characteristic (ROC) Curve (AUC) are used as performance figures [17], \cite{Giannakopoulos2014}. They are all based on the number of long-term frames correctly classified as cough or non-cough. We evaluated two approaches to build the final system: (1) training a single model using information from the three parts of the protocol; (2) training three separate models for each part of the protocol and later combine their outputs using a majority voting scheme, to get the final system output. An additional leave-one-patient-out cross-validation procedure was carried out to assess system generalisation capabilities (see Section \ref{SectionIV.C}).

% SECTION IV: RESULTS
\section{Results} \label{SectionIV}
% SUB-SECTION IV.A: SELECTED FEATURES
\subsection{Selected features} \label{SectionIV.A}
To ensure the generalisation capabilities of the proposed feature selection approach, we carried out the process described in Section \ref{SectionIII.B} for five times. The five randomly selected groups were disjoint sets, so 50\% of observations were employed in this step. Then, five best-29 feature sets were obtained after the combination procedure.

Twenty selected features were common to the five final feature sets: relative power (\(1^{st}, 2^{nd}, 4^{th}\) and \(5^{th}\) frequency band), spectral centroid (\(2^{nd}, 3^{rd}, 4^{th}\) and \(5^{th}\)frequency band), spectral flatness (\(1^{st}, 2^{nd}, 3^{rd}\) and \(4^{th}\) frequency band), spectral roll-off (\(2^{nd}, 3^{rd}\) and \(5^{th}\) frequency band) ratio f50 vs f90 (\(2^{nd}\) frequency band), the spectral entropy, the HR, root MFCC (\(1^{st}\) coefficient), NASE (\(4^{th}\) coefficient). The spectral roll-off from the \(4^{th}\) frequency band, the spectral centroid from the \(1^{st}\) frequency band, the \(11^{th}\) NASE and \(1^{st}\) ASF coefficients were present in four of the five final feature sets.

Finally, the ratio f50 vs f90 (\(3^{rd}\) and \(5^{th}\) frequency band), the relative power (\(3^{rd}\) frequency band), spectral bandwidth (\(2^{nd}\) frequency band) and the \(13^{th}\) NASE coefficient were in three of them.

The final short-term feature space contains these 29 features which were common in at least three of the five trials. Furthermore, it should be pointed out that all the combination procedures stopped in the fifth step or before (see Section \ref{SectionIII.B}), that is, the majority of selected features were common to at least two of the protocol parts and thus robust for different noise levels. Since feature selection was carried out at short-term level, it is worth mentioning that no long-term features from the train or test sets in each of the 5 folds in Fig. 7 was employed for selection.

% SUB-SECTION IV.B: Results for each part of the measurement protocol
\subsection{Main results} \label{SectionIV.B}
Three different models were trained, one for each part of the protocol. All of them were based on the 29 short-term selected features. Table \ref{TableIV} shows the average classification results for each model. The obtained standard deviation was always below 2.98\% and 5.38\% for AvgSD and supervised BoAW, respectively. McNemar's test \cite{Mcnemar1947} was employed to assess statistical significance for SEN and SPE in the comparison between both long term approaches.

\begin{table}
\centering
\caption{Average classification results (\%) for models trained in different parts of the protocol. Statistically significant differences $p<0.05$ (*), $p<0.01$ (**) after applying Mcnemar's test For SEN And SPE.}
\label{TableIV}
	\begin{tabular}{c|c c c| c c c}
		\hline \hline
        & \multicolumn{3}{c|}{AvgSD} & \multicolumn{3}{c}{Supervised BoAW}\\
        \textbf{Part} & \textbf{SEN} & \textbf{SPE} & \textbf{ACC} & \textbf{SEN} & \textbf{SPE} & \textbf{ACC} \\
        \hline
        \(1^{st}\) & 92.71 & 88.58 & 90.65 & 87.70 (*) & 79.86 (**) & 83.78 \\
        \(2^{nd}\) & 88.26 & 88.12 & 88.19 & 81.10 (**) & 81.98 (**) & 81.54 \\
        \(3^{rd}\) & 86.89 & 83.93 & 85.41 & 81.13 (*) & 75.94 (**) & 80.08 \\
        \hline \hline
	\end{tabular}
\end{table}

Performance obtained from AvgSD is higher that the one from supervised BoAW one for all three parts of the protocol with statistical significance for SEN and SPE (note that McNemar's test is not directly applicable to AUC values). Besides, the AvgSD approach is more robust in terms of SEN, since the difference between the ?rst and the third part is smaller (5.82\% vs 6.57\%).

% SUB-SECTION IV.D: Comparison with state-of-the-art
\subsection{Comparison with state-of-the-art} \label{SectionIV.D}
The experimental setup used in Section \ref{SectionIV.B} was used to compare our proposal with three recently proposed cough detectors: 1) the one proposed in \cite{You2017} based on ensembling multiple frequency subband features; 2) our proposal in \cite{Monge2018}, based on moment theory \textit{cepstrogram} characterisation; 3) and the CNN architecture employed by Amoh and Odame in \cite{CNN2016}. The obtained results are presented in Table \ref{TableV}. McNemar's test was again employed to assess statistical significance in the comparisons for SEN and SPE. Fig. 8 shows the protocol-averaged mean ROC curves for all compared methods.

\begin{table}
	\centering
	\caption{Average classification results (\%) obtained with state-of-art methods trained in different parts of the protocol. Statistically significant differences $p<0.05$ (*), $p<0.01$ (**) after applying Mcnemar's test For SEN And SPE.}
	\label{TableV}
	\begin{tabular}{c|c| c c c }
		\hline \hline
		& \textbf{Part} & \textbf{SEN} & \textbf{SPE} & \textbf{ACC} \\
		\hline
		\multirow{3}{*}{\cite{You2017}} & \(1^{st}\) & 77.55 (**) & 76.07 (**) & 76.07 \\
		& \(2^{nd}\) & 75.89 (**) & 73.72 (**)  & 73.72 \\
		& \(3^{rd}\) & 75.89 (**) & 74.92 (**) & 74.93 \\
		\hline
		\multirow{3}{*}{\cite{CNN2016}} & \(1^{st}\) & 74.90 (**) & 89.47 & 89.21 \\
		& \(2^{nd}\) & 74.66 (**) & 90.04 & 89.82 \\
		& \(3^{rd}\) & 69.17 (**) & 87.82 (*) & 87.53 \\
\hline
		\multirow{3}{*}{\cite{Monge2018}} & \(1^{st}\) & 66.36 (**) & 90.28 & 89.80 \\
		& \(2^{nd}\) & 57.38 (**) & 91.27 & 90.73 \\
		& \(3^{rd}\) & 55.01 (**) & 89.79 (*)  & 89.17 \\
		\hline \hline
	\end{tabular}
\end{table}

\begin{figure}
\centering
\includegraphics[width = 8cm]{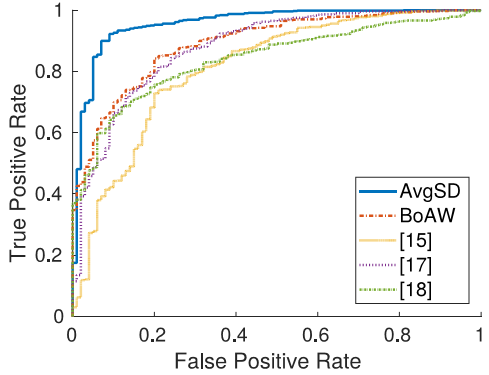}
\caption{Protocol-averaged mean ROC curves for all the compared methods.}
\label{Fig8new}
\end{figure}

The approach in \cite{You2017} shows more robustness among the three methods and offers the highest SEN in the three protocol parts. On the other hand, the moment-based approach \cite{Monge2018} and the CNN \cite{CNN2016} yield higher SPE values although they are not significant at $\alpha =0.05$ level in the first two parts of the protocol. An overall outperformance of our proposal can be seen from AUC values in both tables.

% SUB-SECTION IV.C: LEAVE ONE PATIENT OUT CROSS-VALIDATION
\subsection{Leave-one patient-out cross-validation} \label{SectionIV.C}
\begin{figure*}
\centering
\includegraphics[width = 16cm]{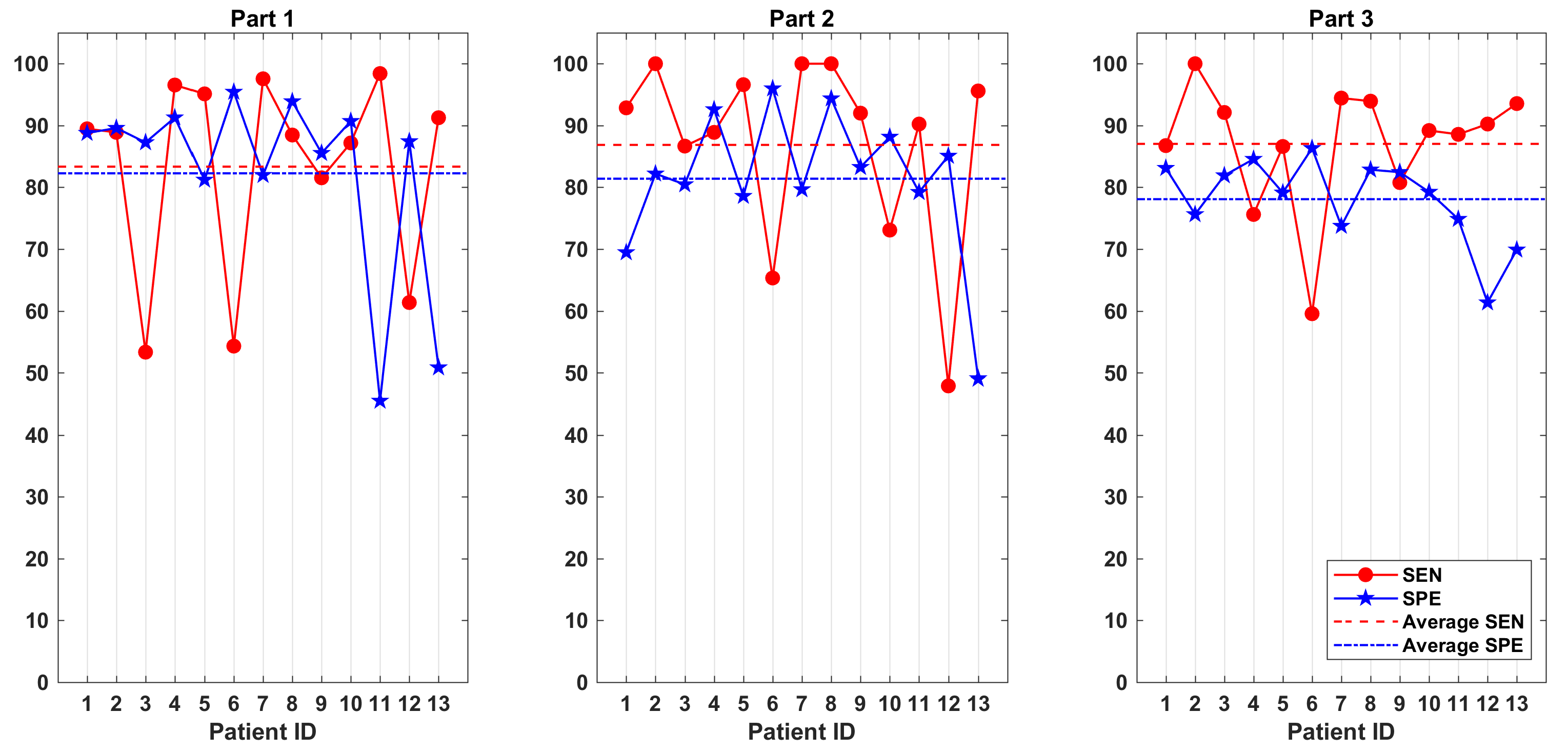}
\caption{Classification results (\%) obtained from leave-one patient-out cross-validation of each part of the acquisition protocol.}
\label{Fig6}
\end{figure*}

In order to assess the generalisation capabilities of our proposal, we performed a second experiment based on a leave-one patient-out cross-validation. In this case, the test set is composed of the whole signal from one patient while the train set is constructed using the signals from the remaining twelve patients. AvgSD was used as the selected high-level representation approach since it performed the best in Section \ref{SectionIV.B}. Fig. \ref{Fig6} shows the obtained classification results in terms of SEN and SPE for each part of the protocol.

Most of the SEN and SPE values in Fig. \ref{Fig6} lay above 80\% even for the \(3^{rd}\) part of the protocol. There are, however, some patients for which the obtained classification performance drops, especially in terms of SEN.

% SUB-SECTION IV.E: Final system
\subsection{Final system performance} \label{SectionIV.E}
\begin{figure*}
\centering
\includegraphics[width = 16cm]{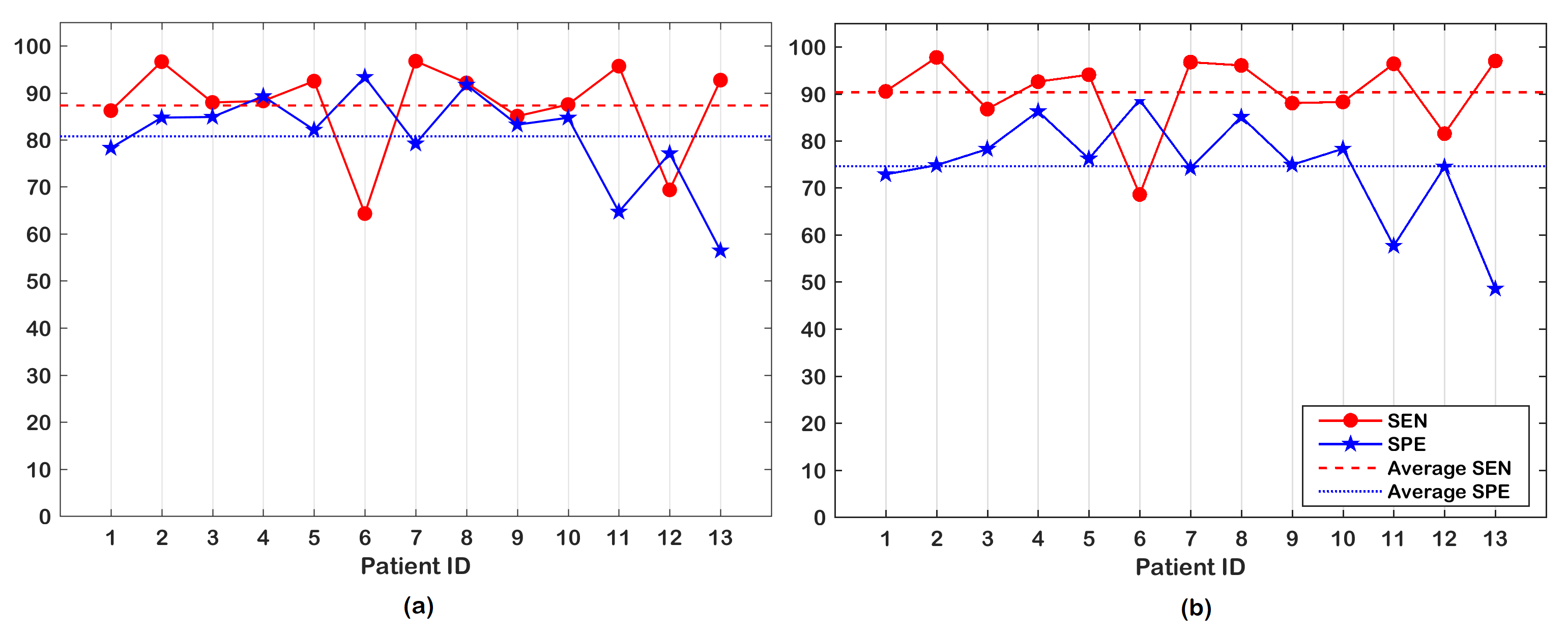}
\caption{Classification results (\%) obtained from leave-one patient-out cross-validation for the final system: (a) when a model is built for each part of the protocol and their outputs are combined using a majority voting scheme (b) when a single model is built using the information from the three parts of the protocol.}
\label{LOPOCV_final}
\end{figure*}

Results provided so far have been obtained from models trained explicitly for each part of the protocol, which present different noise conditions (see Section \ref{SectionII}). In a real scenario, determining the amount of noise in advance to select the specific trained model is not straightforward. As described in Section III-D, we also trained three separate models for each part of the protocol and later combine their outputs using a majority voting scheme, to get the final system output.   The evaluation of the final system is based on a leave-one-patient-out cross-validation strategy as in the previous section. Results are displayed in Fig. \ref{LOPOCV_final}.

\textcolor{black}{The single model approach --Fig. \ref{LOPOCV_final} (b)-- yields higher average SEN than the ensemble one --Fig. \ref{LOPOCV_final} (a)-- at the expense of a drop in SPE. In any case, both systems yield SEN values in the 90\% range and SPE values around 80\%. Moreover, the same trend is observed, being the sixth patient the only one with poor SEN performance in both of them}. There is also a drop in SPE for the last three patients that can be explained by higher noise in the experimental set-up for the third part of the protocol. This can be observed in Fig. \ref{Fig6} (c) for the same patients.

Finally, Fig. 11 shows an illustration of cough events detected and missed by the system, the latter with significantly lower output.

\begin{figure}
\centering
\includegraphics[width = 8.8cm]{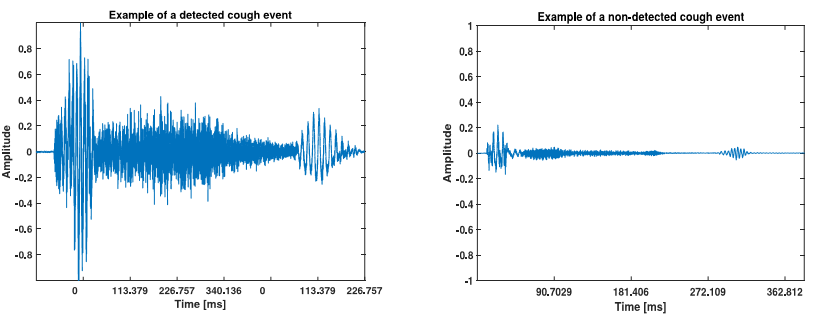}
\caption{Illustration of detected (right) and missed (left) cough events.}
\label{Fig11new}
\end{figure}

\section{Discussion} \label{SectionV}
Our proposal starts from an initial 117-dimension short-term feature set to detect audio cough events in three scenarios: low (part one), moderately (part two) and highly (part three) noisy. After applying the Relieff algorithm and a combination procedure, we identify the twenty-nine most relevant short-term features regardless the environment. These features have been used to build a high-level data representation based on two approaches: AvgSD and supervised BoAW.  \textcolor{black}{The long-term features feed SVM classifiers to get the final classification output}.

The first point to discuss is the followed approach to find the most relevant short-term features regardless the noisy environment. An alternative approach could be to find the best features for each part of the protocol. However, this would make the system sensitive to the noise level. This decision could also lead to two secondary problems. First, the system should have an extra module to identify the type of environment before feature computation. This additional module would potentially reduce the system performance since, if the environment is wrongly detected, the cough characterisation and classification would be suboptimal. This dependency reduces the modularity degree of the system. The second one would be how to recognise the kind of environment.

The final results support the suitability of the proposed feature selection approach. Twenty out of the twenty-nine predictors are present in the five final feature sets.
 Likewise, twenty-three out of the twenty-nine finally selected features are among the features in which we have introduced innovations (separate frequency bands) to adapt their usage for cough segmentation. Thus, our definition of the frequency bands for unidimensional spectral features seems appropriate.

 It is worth noting that the selected spectral short-term features such as band-relative power, and band-specific centroids and flatness, as well as roll-offs have shown meaningful for a number of individual bands. However, these features, when computed globally for the whole band, did not show good performance in [18] for cough detection. Thus, computing them in a band-specific manner has shown their capability to represent cough spectral signatures with noise robustness.  The reason behind the noise robustness of this band-specific feature calculations lies in the way that overlapped noisy sounds affect the signal. From the spectral point of view, noisy background sounds constitute coloured contaminations. Consequently, they modify the signal spectrum locally. By computing these features in distinct frequency bands, such local contamination is avoided (some bands might be affected while others not). Besides, some descriptors such as centroids, crest factor or roll-off are less prone to be distorted by definition \cite{SSCH2006}. Finally, the calculation of these features based on Welch's PSD estimation may also contribute to robustness due to its lower variance compared to other options \cite{Haykin2009}.

Regarding the two high-level approaches, AvgSD was the best-performing despite its simplicity. On the other hand, the dimension of the supervised BoAW feature set is smaller (32 vs 58). In this sense, other values of \(K_{pos}\) and \(K_{neg}\) (e.g., 16 and 32, 32 and 16 or 32 and 32) were tested but AvgSD outperformed it as well. Besides, the standard deviation of the classification results is smaller for AvgSD, so this approach exhibits smaller dependency of the training-test partitions and, consequently, better generalisation properties. In this regard, the use of a simple approach based solely on mean and standard deviation of the short time features has shown good performance compared to the more complex BoAW. More complex approaches using contextual information such as i-vectors \cite{Behravan2015} could also be explored. However, the particular context (continuous, smartphone-based monitoring with low battery consumption) would not benefit from this approach.

The above-mentioned generalisation capability is confirmed by leave-one patient-out cross-validation experiments. Only for one patient (6-th), SEN lies below 80\% in the three parts. The other patients offer good SEN values in at least one of the parts. Therefore, our system is not only robust but also capable of dealing with inter-disease variability \cite{Smith2007} (see Table \ref{TableI}). \textcolor{black}{These results are confirmed when a single model is trained -- Fig. \ref{LOPOCV_final} (b)--. Furthermore, when three models are combined using majority voting -- Fig. \ref{LOPOCV_final} (a) -- higher average SPE is obtained. This behaviour seems plausible since the negative class is much more diverse in terms of types of sounds (see Section \ref{SectionII}), so a single model finds more difficulties at the time of learning this class}.

It is worth mentioning at this point that there exist three patients (6, 11, and 13) for whom the obtained performance is consistently lower. This can be due to several factors. A first conclusion could be extracted from the higher noise profiles presented in some of those patients. For instance, patient 6 shows significant low SNR values in Table II. However, the performance for other patients with low SNR profiles (e.g. patient 9 in part 3 of the protocol) is still good. This leads to the conclusion of a not so good representation in the training group for those patients in the leave-one-patient-out evaluation strategy. This type of problems can be overcome with a larger database to train the system. The size of the study population is actually a limitation of this study. However, the generalization performance of our system is still remarkably good with such a small population.

The approach by You \textit{et al.} \cite{You2017} performed the best among the compared methods. Nevertheless, the pattern recognition capability of our system showed better in the three scenarios. The moment-based approach \cite{Monge2018} and CNN architecture \cite{CNN2016} slightly outperformed our proposal in terms of SPE at the cost of significantly lower SEN figures. Consequently, the associated loss of clinical information (cough patterns) is greater in these systems. Moreover, this experiment confirms the hypothesis that a high-level data representation improves classification performance. The methods in \cite{You2017} and \cite{CNN2016} are short-term approaches whereas the moment-based approach can be understood as a middle point: the short-term observations feed the classifier, but information from adjacent observations is used in the characterisation of each one. It is worth noting that the performance reported in \cite{You2017,Monge2018,CNN2016} for the state-of-the-art methods was higher than the one obtained in our database. This can be explained from a more favourable train-test partition where train and test samples were close in time. In our experiments, the block-wise partition, prevents training samples from being close in time to test ones.

It is also worth mentioning at this point, that our proposal, which is based in craft-engineered features, outperforms the one in \cite{CNN2016}, which relies on modern deep learning approaches based on unsupervised feature extraction. This can be explained from the unbalance between cough and non-cough events. The number of patterns in the positive class might not be enough to train a deep neural network, and thus lower sensitivity values after applying the approach in \cite{CNN2016} to our database can be observed. On the other hand, a pattern recognition engine based on simple features feeding powerful (yet efficient in deployment) classifiers, such as the one here proposed, would allow real-time performance and overcome battery issues in continuous monitoring situations. Deep learning approaches may be too computationally expensive in energy constrained-environments.

Finally, we would like to discuss the clinical applicability of the system. From the medical point of view, cough is not generally a severe symptom, so patients can self-manage their own respiratory diseases \cite{Gibson2002}. If practitioners can rely on an objective cough detector, the number of hospitalisations and consultant referrals from respiratory diseases will be reduced. This would decrease costs for national health systems. Furthermore, this cough monitor is only based on audio recordings so a smartphone- or tablet-based implementation would be easy to deploy \cite{SmartphoneMedDev}. This way, less disruptive patient monitoring could be achieved in real time. Besides, complementary information available from these devices such as location - which can be correlated to pollution and/or pollen levels \cite{CoughPollution}, for instance - or the patients' routine, which can be connected to peaks in the physical activity, could be used with different objectives. These include helping practitioners assess the real impact of cough in the quality of life, treatment follow-up, or extracting the clinical relevance of secondary measures like cough frequency, cough intensity, or cough type (e.g., dry or wet) - which are still undetermined \cite{Birring2015,Wang2017}.

\section{Conclusions} \label{SectionVI}
In this paper, a machine hearing system for robust cough segmentation solely based on audio recordings is proposed. The system characterises cough patterns using twenty-nine short-term features which were selected to be robust in different noisy scenarios. Five frequency bands were defined to adapt the computation of some of these features to the cough spectrum properties. A long-term feature space is generated by using sample statistics over consecutive short-term frames. These feed an ensemble of SVMs, each one trained with samples from different noise scenarios, which provides the final system output after majority voting.

The system is evaluated using a thirteen patient signal database which encompasses three different noisy scenarios. The database is representative of three of the most common respiratory conditions spanning a range of different ages in both men and women. Classification results confirm that our system: (1) outperform so far proposed methods in terms of cough detection, and (2) can cope with three different noisy environments. Furthermore, the system generalisation capability is assessed using a leave one patient out cross-validation strategy to overcome the limitation of having a reduced evaluation dataset. Our system is aligned with a less disruptive and more comfortable patient monitoring, which may benefit patients by enabling self-monitoring of cough symptoms. In addition, our system has potential to provide support in the assessment of treatments and better clinical understanding of cough patterns. Cough audio patterns could be detected and further analysed for this purpose. This could however require a pre-processing step where the effects of noise and other audio events were minimised. Finally, national health systems and economies would also benefit by a reduced number of hospitalisations and productivity loss.

\section*{Acknowledgment}
The authors would like to thank Dr. L. McCloughan, Prof. B. McKinstry, Prof. H. Pinnock, and Dr. R. Rabinovich at the University of Edinburgh for their valuable clinical support. Additional thanks are given to L. Stevenson, D. Bertin, and J. Adams, from Chest Heart and Stroke Scotland, for arranging a patient panel for this research.

\bibliographystyle{IEEEtran2}
\bibliography{TBE18_ref} % \addbibresource{TBE18_ref}

\end{document}